\documentstyle[11pt,cs10,epsf]{article}
 
\begin{document}

\title{Lithium and binarity}

\author{David Barrado y Navascu\'es\altaffilmark{1,2}}
\affil{Harvard--Smithsonian Center for Astrophysics,
    Cambridge, MA 02138}

\altaffiltext{1}{MEC/Fulbright Fellow} 
\altaffiltext{2}{Fellow at the Real Colegio Complutense at Harvard University,
 Cambridge, MA 02138}

\setcounter{footnote}{2}

\begin{abstract}
We present an analysis of the lithium abundances 
 in late spectral type binaries of different ages.
They belong to  several 
open clusters (Pleiades, Hyades and M67), as well as
to Chromospherically Active Binary Systems (CABS). 

All these binaries have reliable ages, since they are members
of well-known open clusters or, in the case of the CABS,
some stellar parameters such masses and radii are accurate
enough to derive ages using theoretical isochrones. Their age span
covers from  barely one hundred million years to several gigayears.

We have compared different stellar properties, such as lithium abundances, 
stellar masses, effective temperatures and orbital periods.
 It is shown that, in general, close
binaries have lithium abundances larger than those characteristic
of single stars or binaries with larger orbital periods. 
The largest difference between the abundances of binaries and single
stars appears at Hyades' age.
The origin of those overabundances are discussed in the context of 
the proposed mechanisms for the lithium depletion phenomenon
and the stellar evolution.

\end{abstract}



\index{*NGC 2682|see M67}
\index{*M 67}

\index{*Hyades}
\index{*Melotte 25|see Hyades}

\index{*Pleiades}
\index{*M 45|see Pleiades}
\index{*Melotte 22|see Pleiades}

\index{*NGC 752}

\index{*UMa Group}

\index{clusters!ages}

\index{lithium}

\index{binary}

\index{*Chromospherically Active Binary Systems|see CABS}
\index{*CABS}

\index{*TLBS}
\index{Tidally Locked Binary Systems|see TLBS}

\section{The problem}

	Lithium depletion depends on different parameters (mass, age, 
metallicity and rotation). The internal structure is very important in
 this phenomenon, but 
all theoretical models predict temperatures at the bottom of the 
convective envelope too low to destroy lithium. Therefore, some extra--mixing 
mechanism, beside pure convection, is needed (see Figure 1).
Nowadays there are a large amount of 
observational data (see the review by Balachandran 1994).
There are at least 2 different mechanisms, related to rotation, which could 
be responsible of the lithium depletion in late spectral--type
 stars. One is related with the
way angular momentum evolves and the mixing associated to it
(rotational breaking), 
and in  other the key factor is the magnetic field and its influence
over the gravitational waves.

\begin{figure}
%
%
\plotone{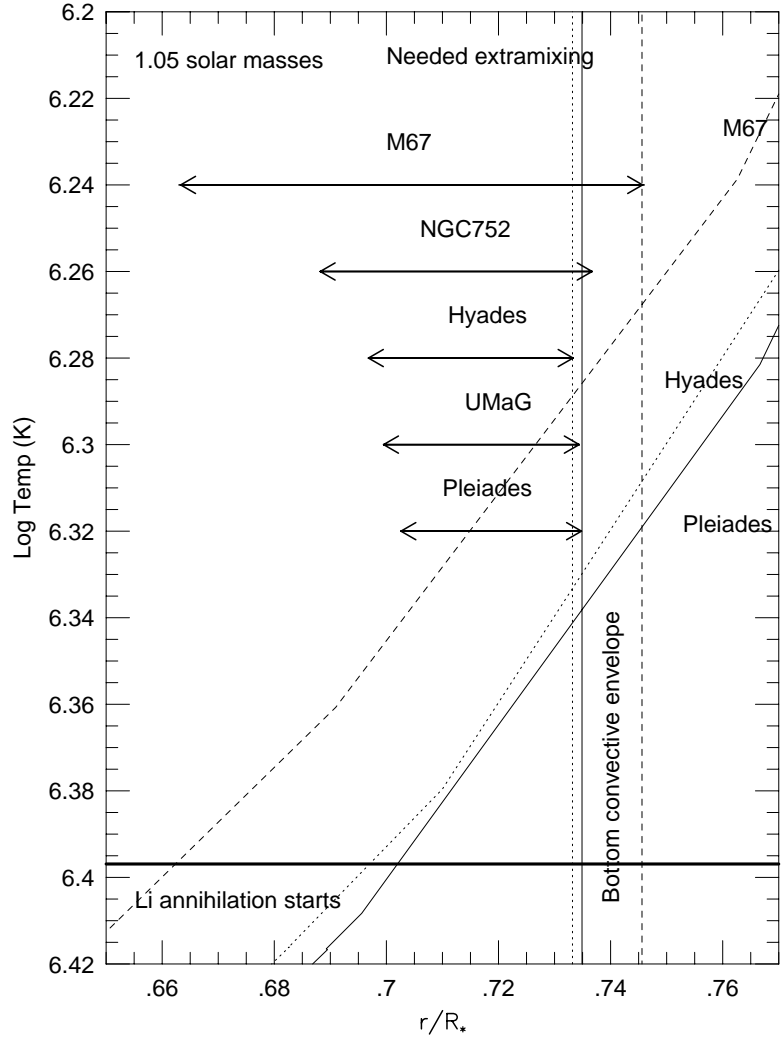}
%
%
\caption{ Temperature against the
fraction of the radii for a 1.05 solar mass star at 
different ages, corresponding to Pleiades, Hyades and M67. The vertical
lines locate the position of the bottom of the convective zone,
 whereas the bold 
horizontal line points out where the lithium is started to be destroyed.
We also indicate the distance between the bottom of the convective envelope 
and the location where the lithium destruction starts.}
\label{fig-1}
\end{figure}

Studies of lithium abundance in binaries can provide interesting 
information about this phenomenon. First, binaries
have different rotational story than single stars
 (either during the first moments of the
evolution in the Premain Sequence --PMS-- 
or afterwards, if they are Tidally Locked Binary Systems --TLBS).
 Second, masses 
 and radii (for eclipsing binaries) are well known, 
allowing a direct contrast between theory and observations. In the case of
eclipsing binaries, 
isochrones can be fitted and ages estimated. Third, in the case of members
of open clusters, we know the age of all members, and it is possible 
to perform direct 
comparisons between single and binary stars of the same mass (looking for
systematic differences) and comparisons with different theories
 can be also carried out.
In this particular context, binaries can help us to understand the role of
rotation in the lithium depletion. In the rotational  braking scenario,
Tidally Locked Binary
Systems would inhibit the lithium depletion due to the transfer of
angular momentum  between the orbit and the rotation, mechanism
 which would be able to 
prevent the mixing of material between the convective envelope
and the  radiative core, where lithium is destroyed
 (Pinsonneault et al. 1989, 1990; Zahn 1994). In the case of a mixing 
mechanism dominated by the presence of magnetic fields and gravitational
waves, lithium
could  be conserved because of the inhibition of the transport of
material due
to gravitational waves (Garc\'{\i}a--L\'opez \& Spruit 1991),
inhibition which would 
appear because  of the presence of strong magnetic  fields 
(Schaztman 1993; Montalb\'an 1994; Montalb\'an \& Schatzman 1996),
due themselves to the rapid rotation  characteristic of TLBS.

\begin{figure}
\
\plottwo{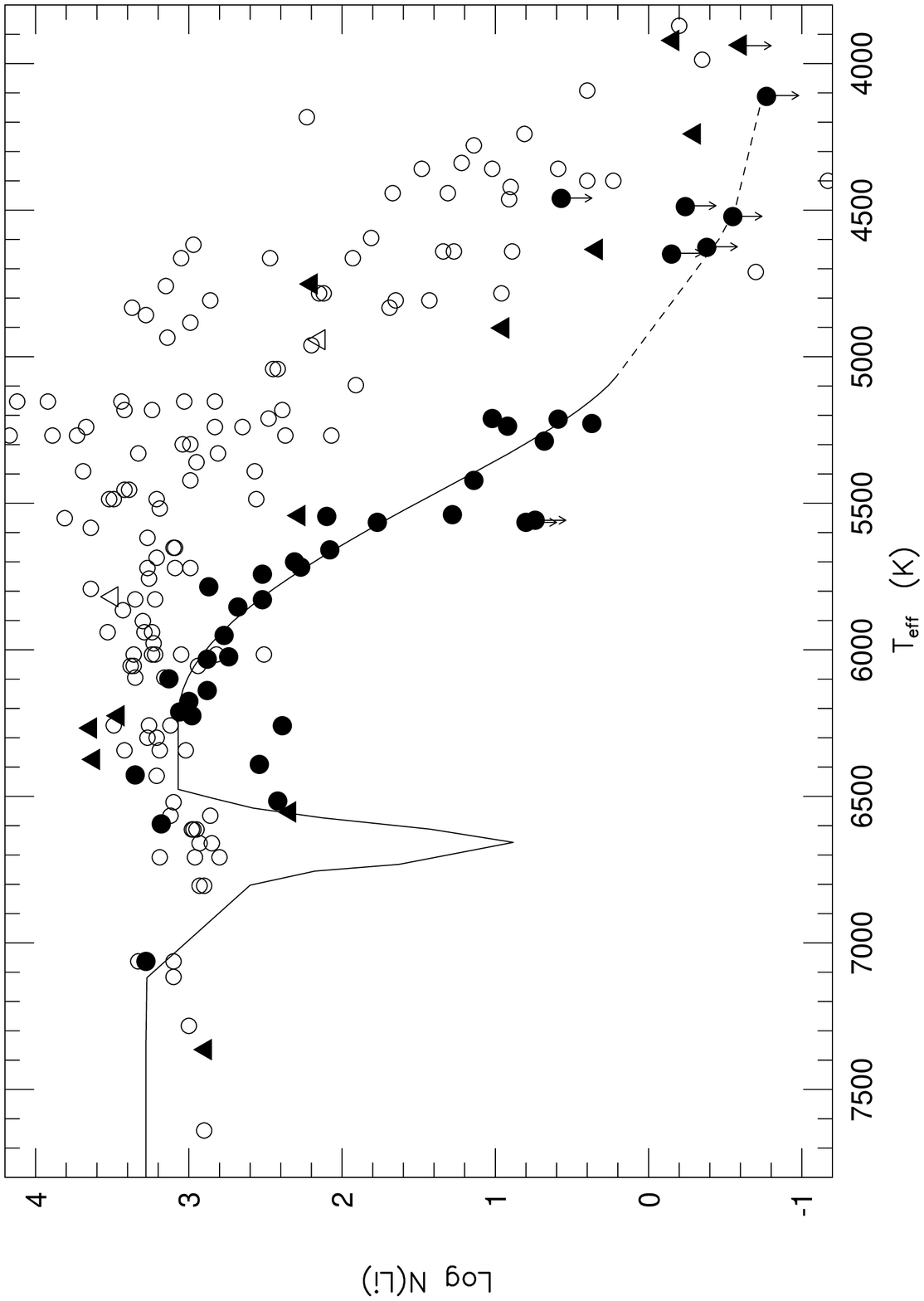}{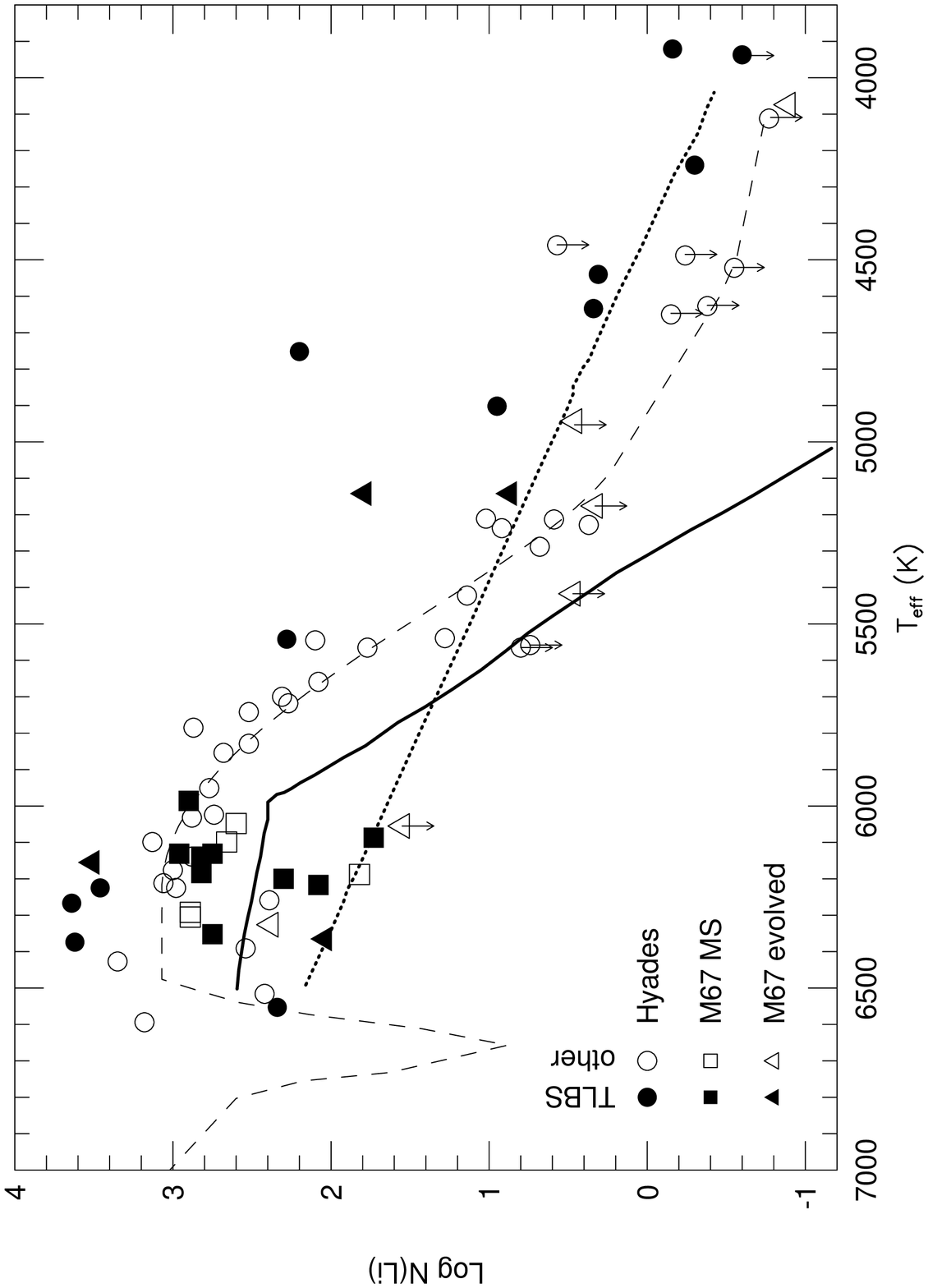}
%
%
\caption{Lithium abundances against effective temperatures.
{\bf a.-} Hyades (solid symbols) and Pleiades (open symbols) data. 
TLBS are superposed as triangles.
{\bf b.-}  Comparison between Hyades main--sequence binaries 
(circles) and M67 binaries (triangles). Solid symbols represent
synchronized 
binaries, whereas open ones show other binaries.
The solid, dashed and dotted lines represent the behavior of main--sequence
M67, evolved
M67 and main--sequence Hyades single stars, respectively.} \label{fig-2}
\end{figure}

	We present here a comprehensive study of the lithium abundances
 of main sequence
 binaries belonging to several open clusters (Pleiades, Hyades and M67), 
having ages ranged from 120 million years to 5 gigayears, as well 
dwarf components of Chromospherically Active Binary Systems.

\section{Lithium, Temperature and Stellar Mass}

The lithium abundances, effective temperatures and other stellar parameters 
analyzed here have been collected from several previous studies. In
 particular, the core of the samples were selected from Soderblom et al.
 (1990) --Pleiades--, Barrado y Navascu\'es \& Stauffer (1996) --Hyades--,
Barrado y Navascu\'es et al. (1997a) --CABS-- and Barrado y Navascu\'es
et al. (1997b) --M67. We have searched through the published literature
to complete these samples (Hobbs \& Pilachowski 1996; Spite et al. 1987;
Butler et al. 1987; Pilachowski \& Hobbs 1987; Garc\'{\i}a--L\'opez et al. 
1988; Boesgaard et al. 1988; Soderblom et al. 1993; Garc\'{\i}a--L\'opez 
et al.  1994; Balachandran 1995).

\begin{figure}
%
%
\plottwo{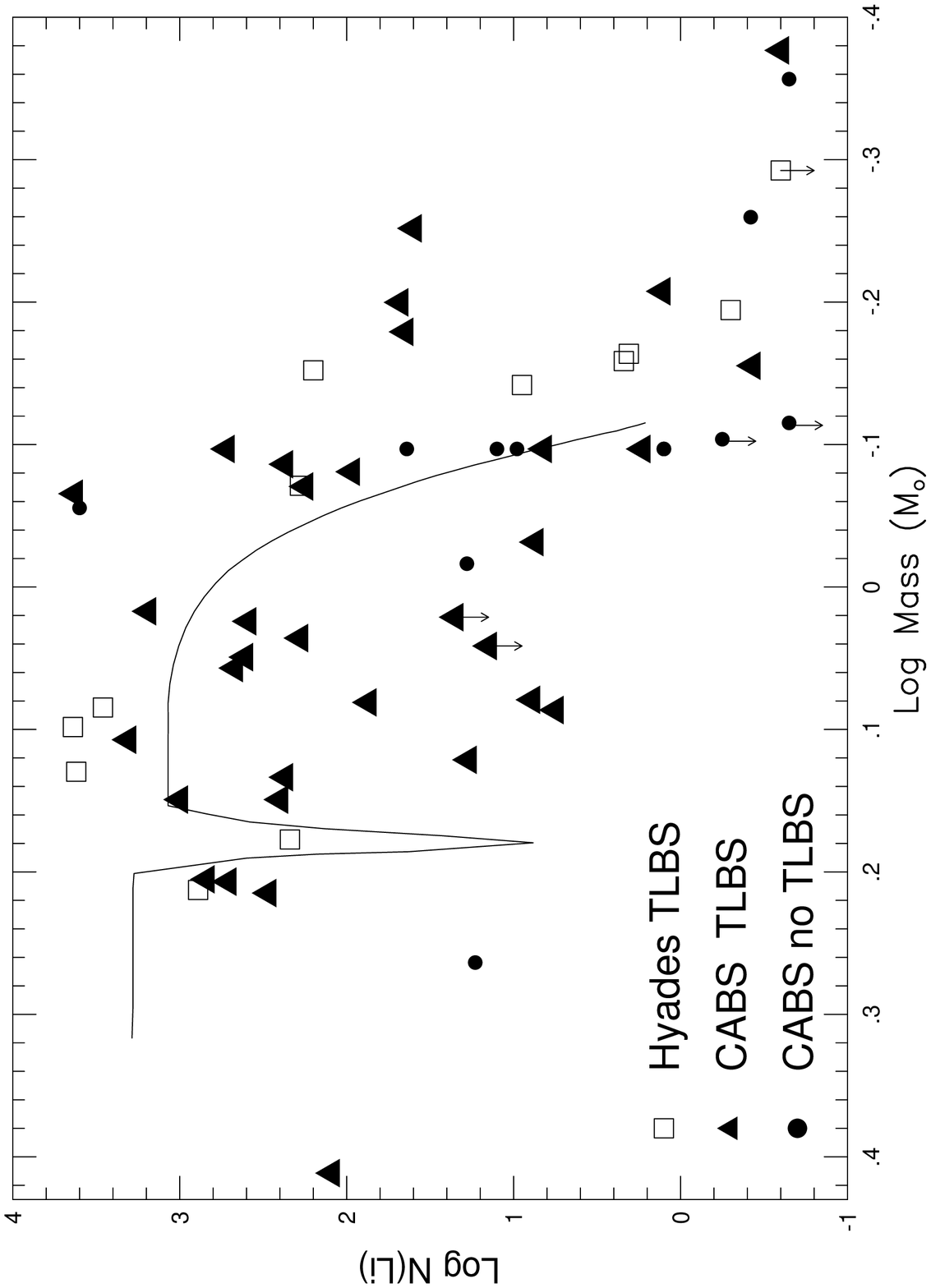}{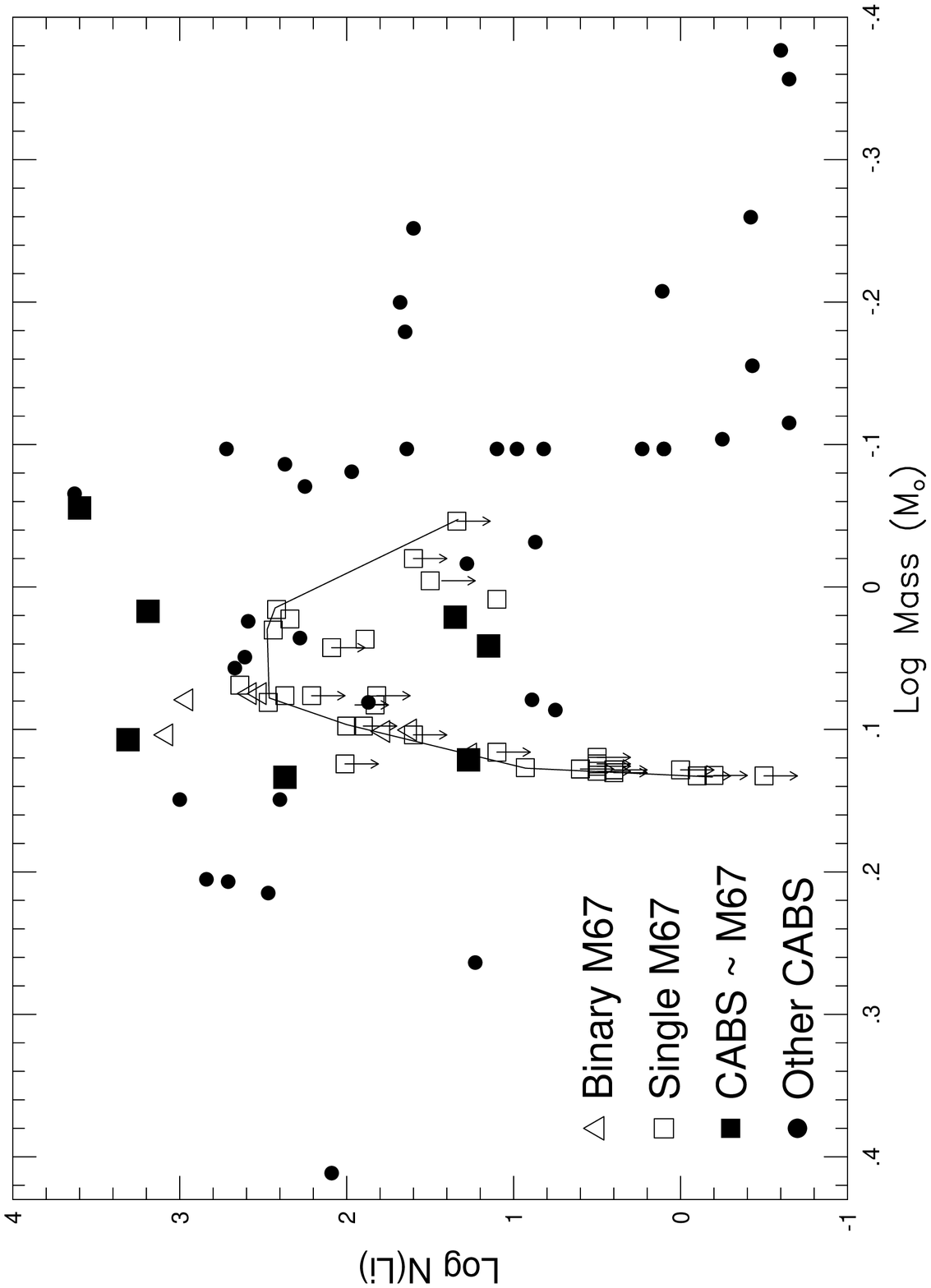}
%
%
\caption{Lithium abundance against stellar mass. Comparison of CABS and binaries
in open clusters of similar age.
{\bf a.-} Hyades binaries and CABS.
{\bf b.-} M67 binaries and CABS.} 
\label{fig-3}
\end{figure}

Essentially, lithium abundances were estimated
 by  measuring equivalent widths, removing the blends with other lines,
 correcting the effect of the companion and using curves of growth to
compute the final value. (See Barrado y Navascu\'es \& Stauffer 1996 
or Barrado y Navascu\'es et al. 1997a for details.
Figure 2a shows Li against T$_{\rm eff}$ for members of Pleiades
 (empty symbols)
 and Hyades (solid symbols). Triangles represent TLBS.
Note that all Hyades stars are binaries in the figure (the behavior of 
single  stars is shown by a solid--dashed line), whereas open circles are
single and wide binaries from Pleiades simultaneously.
 It is clear that Hyades TLBS
have lithium excesses above the average value of other Hyades single and
 binary stars. This does not happen in Pleiades
(note that there a few data available in this case).
 Therefore, it seems that the inhibition of the
lithium depletion takes places after an age of $\sim$100 Myr.

Figure 2b   represents M67 data together the Hyades. 
Solid and empty symbols are
TLBS and other binaries, respectively. Hyades are shown as circles. Main
sequence and evolved M67 binaries appear as squares and triangles,
 respectively. The long dashed line is the average lithium abundance of single
Hyades stars. The solid and short dashed lines correspond to the average
abundance of single MS and giant M67 members, respectively. Although
some M67 TLBS have lithium excesses, it is easy to notice that several 
main sequence M67
TLBS have undergone an important depletion, contrary to what happens
in the Hyades.

The comparison between Hyades and M67 main  sequence G 
  binaries shows that their
rates of lithium depletion are different, in the sense that the
 average abundance of Hyades TLBS is 0.6 dex higher
 than the average abundance of other binaries or respect
single stars belonging to the same cluster, whereas main sequence 
G binaries of
M67 have an average abundance  only 0.22 dex higher than that 
corresponding to the single main sequence G stars.
The difference between the abundances of binaries of Hyades and M67 is, 
on average, 0.65 dex, whereas the value is 0.37 in the case of the
single stars. These facts seem to indicate that
essentially the process which produced the overabundances,
at least in the case of main--sequence stars in the range 
6500$>$ T$_{\rm eff}$ $>$ 5500 K,  took place before  having the
Hyades' age. (It could last a little bit longer.) Moreover, 
since the difference between the average abundances of binaries and 
single stars are larger in Hyades than in M67, it might be that not only
 the inhibition in the lithium depletion was stopped
at an intermediate age, but that effect
of binarity  has been reversed. That is, the depletion could be
faster in binaries than in single stars during the lapse of
time passed since   M67 had Hyades' age to the present. Note
that in this whole  discussion, we have assumed that the properties
of M67 members where analogous to the Hyades stars. In particular, 
that the lithium abundance distribution was similar when M67 was 
600--800 Myr old.

\begin{figure}
%
%
\plotone{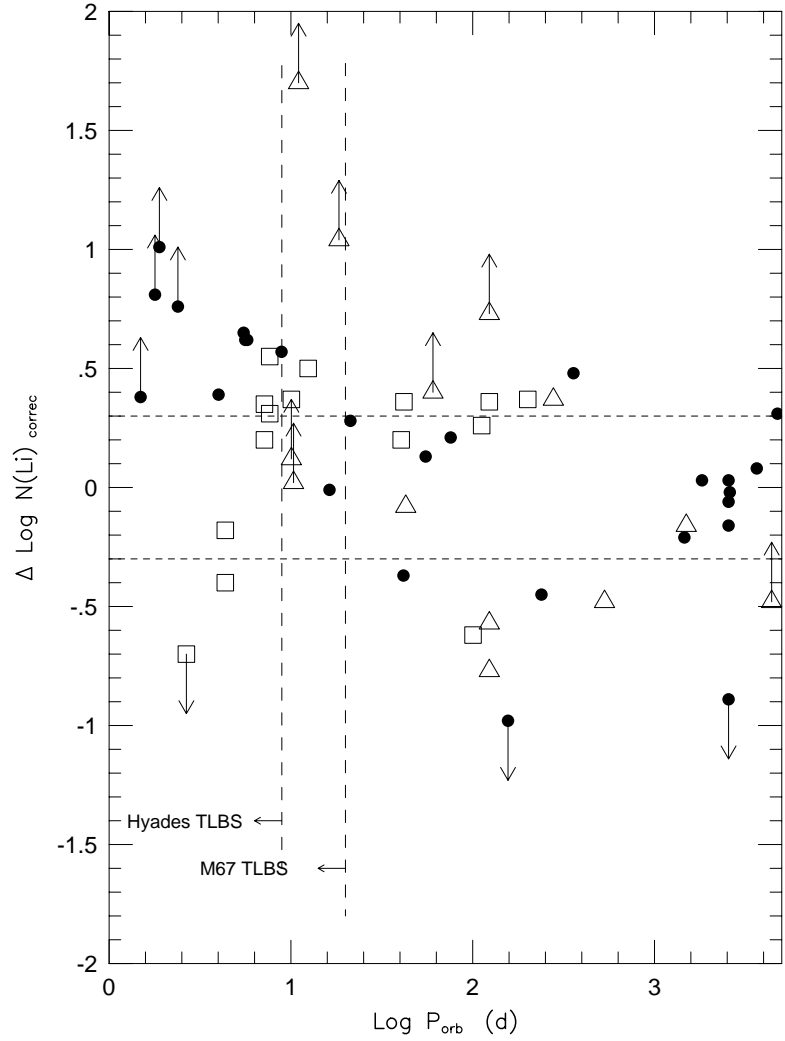}
%
%
\caption{ Lithium excesses against orbital periods. 
 Main--sequence Hyades binaries are
shown as solid circles. Main Sequence and evolved M67 binaries
appear as
open squares and triangles. The vertical lines differentiate the
systems 
which have synchronization between the orbital and the rotational
periods for both clusters.} \label{fig-4}
\end{figure}

Additional information about how lithium is depleted in old MS binaries can be 
obtained by studying Chromospherically Active Binary Systems.
In order to facilitate the interpretation of the data, we substituted
 T$_{\rm eff}$ 
by the stellar mass in Figure 3a.  Hyades TLBS are shown as open squares,
CABS with similar age (see Barrado et
al. 1994 for the estimation of ages) to Hyades' appear as solid squares, 
whereas  other CABS are shown as solid circles (all of them, when the age is
known, are older). CABS have a large spread in the abundance for a mass
close to 0.8 M$_\odot$. Several CABS have also larger abundances than similar
 Hyades stars, although the former group is older. 
Figure 3b is as Figure 3a, 
but in this case M67 binaries are shown instead Hyades
members (see key in the figure).  Despite the fact that
CABS have large uncertainties in their abundances and in the other stellar
parameters (except, in the case of eclipsing binaries, for masses and radii), 
this plot adds new evidences to the connection between lithium depletion and 
binarity. Similar results have been found in evolved CABS by Pallavicini
et al. (1992, 1993) and Randich et al. (1994).

\begin{figure}
%
%
\plotone{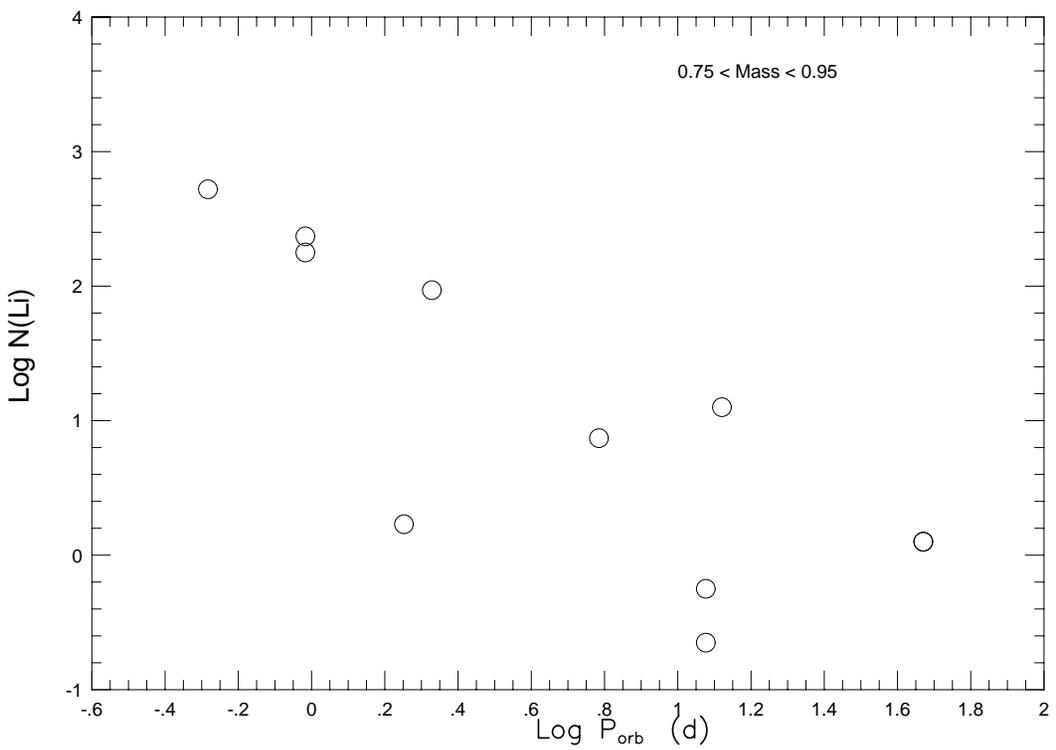}
%
%
\caption{ Li abundances against orbital periods for CABS.} \label{fig-5}
\end{figure}

\section{Orbital Period and Lithium abundance}

Barrado y Navascu\'es \& Stauffer (1995) obtained that {\it all}
TLBS in the Hyades have clear lithium excesses and that these are related
with the orbital period, as shown in Figure 4. An analogous result has
been obtained by Barrado y Navascu\'es et al. (1997b) for M67.
Figure 4 shows the differences of lithium abundance between the estimated 
value for binaries and the computed average of stars at the same color
against the orbital period. Hyades binaries are represented with solid
circles, M67 dwarfs appear as empty squares and evolved M67 binaries
are plotted as empty triangles. Note that the cut--off of the orbital periods
for Hyades in order to have synchronization between the orbit and the 
rotation  is smaller than in  the case of M67, due to the age difference.
 In the case of M67, not all TLBS have overabundances. The situation is 
similar for CABS (Figure 5).
 In this case, we have shown the lithium abundances
against the orbital periods for the subsample of dwarf components of CABS
having masses in the interval 0.75--0.95 M$_\odot$ (these stars
have similar age). There is a clear trend, although, as happens in the 
case of M67, short orbital period TLBS can have low abundances.

Together with the interpretation provided in the previous section, 
it would be also possible to speculate that synchronization is
not enough 
for old low mass stars. On the other hand, if the story of the
rotation
is the most important factor (that is, the way the angular
momentum
has been transported from the orbit to the rotation), it would be
possible
to find systems with synchronization and low abundances. It
would be due
to the fact that initial rotational periods would be very
important in that
 case. Stellar  rotation can change dramaticly
during the 
first stages of the synchronization, with sudden increases and
decreases,
 depending on the particular combination of orbital and rotational
periods, and
stellar masses. Therefore, a very enhanced mixed of material can
take place
during some of these first moments, leading to rapid lithium depletion.
After 
synchronization, the lithium depletion would be inhibited. 
Although the rotational braking mechanism seems to fit better the 
observations, the presence gravitational waves as a way to
mix material cannot be ruled out yet. A detailed calculation could provide 
some quantitative arguments in order to select the proper mechanism.

\section{Conclusions}

There is a clear connection between the way lithium is depleted
and binarity. We have shown that some binaries have larger abundances
than similar single stars.
Therefore, rotation is an essential parameter in the 	
lithium depletion phenomenon. The mechanism scenario and the role of binarity
are not clear.

The comparison of the lithium abundances of binaries in several open clusters
shows that the lithium excesses are maximum at Hyades' age.
Binaries at the Pleiades's age have lithium abundances compatible
with single stars.

 In the case of  main sequence 
F--K binaries at Hyades' age, lithium depletion is inhibited 
always if P$_{\rm orb}\le$8 days (if the system arrives synchronized at the
 MS).

 Older TLBS exhibits, but not always, lithium excesses. Their nature
of these excesses is not clear, and they could be a remain of those
excesses developed during the first hundred years in the MS.

\acknowledgments
DBN acknowledges the support by Real Colegio Complutense at Harvard University,
 and the   Commision for Cultural,
Educational and Scientific Exchange between the United States of
America and Spain.

\end{document}